\documentclass[11pt]{iopart}

\usepackage{amsbsy,amssymb,mathrsfs}

\newcommand{\al}{\alpha}
\newcommand{\be}{\beta}
\newcommand{\ga}{\gamma}
\newcommand{\de}{\delta}
\newcommand{\m}{\mu}
\newcommand{\n}{\nu}

\newcommand{\si}{\sigma}
\newcommand{\eps}{\epsilon}

\newcommand{\Si}{\Sigma}
\newcommand{\ta}{\tau}
\newcommand{\ups}{\upsilon}
\newcommand{\lam}{\lambda}

\newcommand{\om}{\omega}
\newcommand{\mc}{\mathcal}
\newcommand{\ce}{\mc{E}}
\newcommand{\cm}{\mc{M}}
\newcommand{\cs}{N}
\newcommand{\cl}{\mc{L}}

\newcommand{\ca}{\mc{A}}

\newcommand{\na}{\nabla}

\newcommand{\ch}{\mc{H}}
\newcommand{\cq}{\mc{Q}}
\newcommand{\cg}{\mc{G}}
\newcommand{\cf}{\mc{F}}
\newcommand{\en}{n}
\newcommand{\nn}{\nonumber}
\newcommand{\ex}{\mc{C}}

\newcommand{\sch}{{\mbox{s}}}
\newcommand{\veh}{{\mbox{v}}}
\newcommand{\teh}{{\mbox{\tiny{T}}}}
\newcommand{\dst}{W}

\begin{document}

\title[1+1+2 gravitational perturbations on LRS class II space-times]{1+1+2 gravitational perturbations on LRS class II space-times: Decoupling GEM tensor harmonic amplitudes}

\author{R. B. Burston}

\address{Max Planck Institute for Solar System Research,
37191 Katlenburg-Lindau, Germany}
\eads{\mailto{burston@mps.mpg.de}}

\begin{abstract}
This paper considers gauge-invariant and covariant gravitational perturbations on arbitrary vacuum {\it locally rotationally symmetric} (LRS) class II space-times. Ultimately, we derive four decoupled equations governing four specific combinations of the {\it gravito-electromagnetic} (GEM) 2-tensor harmonic amplitudes. We use the gauge-invariant and covariant 1+1+2 formalism which Clarkson and Barrett \cite{Clarkson2003} developed for analysis of vacuum Schwarzschild perturbations. In particular we focus on the first-order 1+1+2 GEM system and use linear algebra techniques suitable for exploiting its structure. Consequently, we express the GEM system new 1+1+2 complex form by choosing new complex GEM tensors, which is conducive to decoupling. We then show how to derive a gauge-invariant and covariant decoupled equation governing a newly defined complex GEM 2-tensor.  Finally, the GEM 2-tensor is expanded in terms of arbitrary tensor harmonics and linear algebra is used once again to decouple the system further into 4 real decoupled equations. 
\end{abstract}

\maketitle

\section{Introduction}

The gauge-invariant and covariant 1+1+2 formalism was first developed by Clarkson and Barrett \cite{Clarkson2003} for an analysis of vacuum gravitational perturbations to a covariant Schwarzschild space-time.  This was further developed in \cite{Betschart2004} who considered both scalar and electromagnetic (EM) perturbations to arbitrary {\it locally rotationally symmetric} (LRS) class II space-times \cite{Ellis1967,Stewart1968,Elst1996}, where they were able to derive generalized Regge-Wheeler \cite{Regge1957} (RW) equations governing the 1+1+2 EM scalars, $\mathscr{E}$ and $\mathscr{B}$. Subsequent to this, we also considered EM perturbations to LRS class II space-times \cite{Burston2007EMBP,Burston2007EMVH}. Therein, we used linear algebra techniques to show that the first-order 1+1+2 Maxwell's equations naturally decouple by choosing new complex variables. Consequently, we expressed Maxwell's equations in a new 1+1+2 complex form that is suited to decoupling. We reproduced the generalized RW result in a new complex form and further established that the EM 2-vectors, $\mathscr{E}_\m$ and $\mathscr{B}_\m$, also decouple from the EM scalars. The EM 2-vectors were expanded into two polar perturbations $ \{\mathscr{E}_\veh, \bar  \mathscr{B}_\veh\}$ and two axial perturbations $ \{\bar \mathscr{E}_\veh,   \mathscr{B}_\veh\}$ using the arbitrary vector harmonic expansion developed in \cite{Clarkson2003,Betschart2004}. Finally, we once again used linear algebra techniques, and derived four {\it real} decoupled equations governing the four combinations of the 2-vector harmonic amplitudes \cite{Burston2007EMVH}. The precise combinations which decoupled were found to be, for the polar perturbations $ \{\mathscr{E}_\veh- \bar  \mathscr{B}_\veh,  \mathscr{E}_\veh+\bar  \mathscr{B}_\veh\}$, and for the axial perturbations $ \{\bar \mathscr{E}_\veh-   \mathscr{B}_\veh, \bar \mathscr{E}_\veh+   \mathscr{B}_\veh\}$.

 In this paper, we consider both gravitational and energy-momentum perturbations to arbitrary vacuum LRS class II space-times using the 1+1+2 formalism. The primary focus is with the first-order GEM system as it is well established to have remarkably similar mathematical structure to Maxwell's equations \cite{Bel1958,Maartens1998}. We use similar techniques as in \cite{Burston2007EMVH}, which was successful to fully decouple the EM 2-vector harmonic components, and ultimately show that this is also successful for fully decoupling the GEM 2-tensor harmonic components.
 
 In Section \ref{PreviousWork} we collate the important results arising from Clarkson and Barrett's 1+1+2 formalism and the background LRS class II space-time is reproduced from \cite{Betschart2004}. Also, the scalar and 2-vector  harmonic expansion formalism is taken from \cite{Betschart2004}, and we provide a new generalization of the {\it spherical} tensor harmonics developed in \cite{Clarkson2003} to tensor harmonics.  We use precisely the same notation as in \cite{Clarkson2003, Betschart2004, Burston2007EMVH} as well as introduce some new quantities which are well defined throughout. In Section \ref{Grperts}, we carefully define the first-order perturbations (including the energy-momentum quantities) to be gauge-invariant according to the Sachs-Stewart-Walker Lemma \cite{Sachs1964,Stewart1974},  We proceed to write the first-order GEM system, conservation equations and the Ricci identities. In Section \ref{decouplingphimn} we derive the decoupled equations and consider tensor harmonic expansions.

\section{Preliminaries}\label{PreviousWork}

The purpose of this section is to present the necessary results for the current series of papers on gravitational and energy-momentum perturbations to arbitrary vacuum LRS class II space-times.

\subsection{Clarkson and Barrett's 1+1+2 formalism}

The 1+3 formalism is very well established (see for example \cite{Ellis1967,Bel1958,Ehlers1993}) whereby, a four-velocity $u^\m$ is defined such that it is both time-like and normalized ($u^\al u_\al=-1$). Consequently, all quantities and governing equations are decomposed by projecting onto a 3-sheet which is orthogonal to $u^\m$, and hence they are called 3-tensors, and in the time-like direction. The essential ingredient for Clarkson and Barrett's 1+1+2 formalism is to further decompose the 1+3 formalism by introducing a new ``radial" vector $\en^\m$ which is space-like and normalized ($\en^\al \en_\al=1$) and orthogonal to $u^\m$. In this way, all 3-tensors may be further decomposed into 2-tensors which have been projected onto the 2-sheet orthogonal to both $\en^\m$ and $u^\m$ and in the radial direction.  The  covariant derivative of the four-velocity in standard 1+3 notation is 
\begin{eqnarray}
\na_\m u_\n &=& \si_{\m\n} +\frac 13\,\theta \,h_{\m\n}- u_\m \dot u_\n +\epsilon_{\m\n\al}\om^\al,
\end{eqnarray}
where $\na_\m$ is the covariant derivative operator, $\si_{\m\n}$ and $\theta$ are the shear and expansion of the 3-sheets, $h_{\m\n}$ is a tensor that projects onto the 3-sheets, $\eps_{\m\n\si}$ is the Levi-Civita 3-tensor and $\om^\m$ is the vorticity.  Finally, the acceleration vector is $\dot u^\m$ where the ``dot" derivative is defined $\dot X_{\m\dots\n} := u^\al \na_\al X_{\m\dots\n}$ and $X_{\m\dots\n}$ represents any quantity. Clarkson and Barrett irreducibly split these standard 1+3 quantities into 1+1+2 form according to
\begin{eqnarray}
\dot u_\m &=&\ca\, \en_\m +\ca_\m,\\
\om_\m &=&\Omega\, \en_\m + \Omega_\m ,\\
\si_{\m\n} &=& \Si_{\m\n}-\frac 12\,\cs_{\m\n}\,\Si +2\, \Si_{(\m}\, \en_{\n)} +\Si\, \en_\m \, \en_\n.
\end{eqnarray}
The 1+3 GEM fields are also decomposed,
\begin{eqnarray}
E_{\m\n} &=& \ce_{\m\n}-\frac 12\,\cs_{\m\n}\,\ce +2\, \ce_{(\m}\, \en_{\n)} +\ce\, \en_\m \, \en_\n,\\
H_{\m\n} &=& \ch_{\m\n}-\frac 12\,\cs_{\m\n}\,\ch +2\, \ch_{(\m}\, \en_{\n)} +\ch\, \en_\m \, \en_\n,
\end{eqnarray}
where $E_{\m\n}$ and $H_{\m\n}$ are respectively the electric and magnetic parts of the Weyl tensor, $C_{\m\n\si\ta}$. In a similar fashion, the 3-covariant derivative ($D_\m$) of the radial vector is decomposed into 1+1+2 form according to
\begin{eqnarray}
D_\m \en_\n &=& \en_\m \,a_\n +\frac 12\,\phi\,\cs_{\m\n} +\xi\,\epsilon_{\m\n} +\zeta_{\m\n},
\end{eqnarray}
where $\zeta_{\m\n}$ and $\phi$ are respectively the shear and expansion of the 2-sheets, $\cs_{\m\n}$ is a tensor that projects onto the 2-sheets, $\xi$ represents the twisting of the sheet and $\eps_{\m\n}$ is the Levi-Civita 2-tensor. Also, the acceleration 2-vector is $a_\m := \hat n_\m$ where the ``hat" derivative is defined $\hat W_{\m\dots\n} := n^\al D_\al W_{\m\dots\n}$ and $W_{\m\dots\n}$ represents a 3-tensor. Finally, the ``dot" derivative of the radial normal is also split according to
\begin{eqnarray}
\dot \en_\m &=& \ca\, u_\m +\al_\m.
\end{eqnarray}
Therefore, the irreducible set of 1+1+2 quantities, and in accord with standard terminology, is
\begin{eqnarray}
\mbox{scalars:}     &\{\ca, \phi,\Si,\theta,\ce, \ch,\Lambda,\xi,\Omega\},&\nn\\
\mbox{2-vectors:} &\{a^\m,\al^\m,\Omega^\m,\ca^\m,\Si^\m,\ce^\m,\ch^\m\},&\nn\\
\mbox{2-tensors:}\,\,\,  &\{\Si_{\m\n},\zeta_{\m\n},\ce_{\m\n},\ch_{\m\n} \},&
\end{eqnarray}
where the cosmological constant ($\Lambda $) has also been included. Furthermore, the energy-momentum quantities, heat-flux and anisotropic pressure,  become respectively \cite{Betschart2004},
\begin{eqnarray}
q^\m &=& \cq\,\en^\m+\cq^\m ,\\
\pi_{\m\n} &=& \Pi_{\m\n}-\frac 12\,\cs_{\m\n}\,\Pi +2\, \Pi_{(\m}\, \en_{\n)} +\Pi\, \en_\m \, \en_\n.
\end{eqnarray}
Thus, the irreducible 1+1+2 energy-momentum quantities are
\begin{eqnarray}
\fl \mbox{scalars:}  \,\,   \{\mu, p, \cq,\Pi\},\qquad \mbox{2-vectors:}\,\, \{\cq^\m , \Pi^{\m}\} \qquad\mbox{and}\qquad \mbox{2-tensor:}\,\,\{ \Pi_{\m\n} \}.
\end{eqnarray}
where $\m$ is the mass-energy density  and $p$ is the isotropic pressure.

\subsection{Background Vacuum LRS class II space-time}\label{backgroundlrsII}

The background comprises the most general vacuum LRS class II space-time and is defined by six non-vanishing LRS class II scalars 
\begin{eqnarray}
\mbox{LRS class II}:\{\ca, \phi,\Si,\theta,\ce, \Lambda\}.
\end{eqnarray}
The background Ricci identities for both $u^\m$ and $\en^\m$ and the Bianchi identities yields a set of evolution and propagation equations governing these scalars. They were first presented in \cite{Clarkson2003} for a covariant Schwarzschild space-time and generalized to non-vacuum LRS class II space-times in \cite{Betschart2004} for which  we reproduce them here for the vacuum case,
\begin{eqnarray}
\Bigl(\cl_\en+\frac 12\,\phi\Bigr) \phi+\Bigl(\Si-\frac 23\,\theta\Bigr)\Bigl(\Si+\frac 13\,\theta\Bigr)+\ce+\frac 23\,\Lambda=0, \label{rb1}\\
\Bigl( \cl_\en +\frac 32\,\phi\Bigr) \Si -\frac23\,\cl_\en \theta=0,\label{rb2}\\
\Bigl( \cl_\en +\frac 32\,\phi\Bigr)\ce=0, \label{rb3}\\
\Bigl(\cl_u -\frac 12\,\Si+\frac 13\,\theta\Bigr) \phi+\ca\,\Bigl(\Si-\frac 23\, \theta\Bigr)=0,\\
\Bigl(\cl_u-\frac 12\,\Si +\frac 13\,\theta\Bigr)\Bigl(\Si -\frac23\,\theta\Bigr)+\ca\,\phi+\ce+\frac 23\,\Lambda= 0,\\
\Bigl(\cl_u  -\frac 32\, \Si+\theta\Bigr)\ce =0,\\
\Bigl(\cl_\en +\ca-\frac 12\,\phi\Bigr)\ca-\frac 32\,\Bigl(\cl_u+\frac 12\,\Si +\frac 23\,\theta\Bigr)\Si -\frac 32\,\ce+\Lambda=0,\\
\Bigl(\cl_u+\Si+\frac 13\,\theta\Bigr)\Bigl(\Si+\frac 13\,\theta\Bigr) -(\cl_\en+\ca)\ca+\ce,\label{rb8}\\
\de_\m \ce =\de_\m \phi =\de_\m \ca =\de_\m\theta=\de_\m\Si =0,\label{rb9}
\end{eqnarray}
where $\de_\m$ is the covariant 2-derivative associated with the 2-sheet. Moreover, in addition to the ``dot" and ``hat" derivatives, we will also use the Lie derivative, $\cl_u$ and $\cl_\en$, (where, for example, the standard definition can be found in \cite{dinverno}). This allows us to neatly express the equations using covariant differential operators. Since the system \eref{rb1}-\eref{rb9} considers only scalars, they simply become usual directional derivatives in this case and are equivalent to the ``dot" and ``hat" derivatives,
\begin{eqnarray}
\cl_u \psi = \dot \psi \qquad \mbox{and}\qquad \cl_\en \psi = \hat \psi.
\end{eqnarray}
Furthermore, for a 2-vector $\psi_{\m} $ and 2-tensor $\psi_{\m\n}$, they are related as follows,
\begin{eqnarray}
\Bigl( \cl_\en-\frac 12 \,\phi\Bigr) \psi_{\bar\m} = \hat\psi_{\bar\m}\qquad&\mbox{and}&\qquad  (\cl_\en-\phi) \psi_{\bar\m\bar\n} = \hat \psi_{\bar\m\bar\n},\\
\Bigl(\cl_u+\frac 12\, \Si -\frac 13\,\theta\Bigr) \psi_{\bar\m} =\dot{\psi}_{\bar\m}& \qquad\mbox{and}&\qquad\Bigr(\cl_u+ \Si -\frac 23 \,\theta\Bigr)\psi_{\bar\m\bar\n} =\dot{\psi}_{\bar\m\bar\n}.
\end{eqnarray}
It is also convenient to introduce five more definitions for the 2-gradients of the LRS class II scalars that arise in \eref{rb9}. Three of these arise in  \cite{Clarkson2003}, 
\begin{eqnarray}
X_\m :=  \de_\m \ce,\qquad Y_\m :=  \de_\m \phi \qquad\mbox{and}  \qquad Z_\m :=  \de_\m \ca, \label{xysa}
\end{eqnarray}
and two new definitions are made to account  for the additional complications of an {\it arbitrary} LRS class II background,
\begin{eqnarray}
V_\m := \de_\m\Bigl(\Si+\frac 13\,\theta\Bigr) \qquad\mbox{and}\qquad \dst_\m &:=&\de_\m\Bigl(\Si-\frac 23\,\theta\Bigr).
\end{eqnarray}

Finally, as in \cite{Betschart2004} we also find it useful to work with the extrinsic curvature and it also comes with evolution and propagation equations,
\begin{eqnarray}
K=\frac 14\,\phi^2 -\frac 14\Bigl(\Si-\frac 23\,\theta\Bigr)^2 -\ce +\frac 13\,\Lambda,\\
(\cl_\en+\phi)K=0\qquad\mbox{and}\qquad \Bigl(\cl_u-\Si+\frac 23\,\theta\Bigr) K =0.
\end{eqnarray}

\subsection{Harmonic Expansions}

The {\it spherical harmonic expansions} for 1+1+2 scalars, 2-vectors and 2-tensors were first presented in \cite{Clarkson2003} for the specific Schwarzschild case. This was subsequently generalized to {\it harmonic expansions} for both scalars and 2-vectors in \cite{Betschart2004}. In this section, we reproduce the necessary results from \cite{Betschart2004} as well as include a new generalization of the 2-tensor {\it spherical } harmonics in \cite{Clarkson2003} to 2-tensor harmonics. Dimensionless sheet harmonic functions $Q$ (defined on the background) are defined
\begin{eqnarray}
\de^2 Q : =-\frac{k^2}{r^2}\, Q\qquad\mbox{and} \qquad \hat Q= \dot Q =0,
\end{eqnarray}
where $k^2$ is real and the 2-Laplacian is defined $\de^2 := \de^\al \de_\al$. The scalar function $r$ is defined by the following covariant equations
\begin{eqnarray}
\Bigl(\cl_\en-\frac12\,\phi\Bigr) \,r =0,\qquad \Bigl(\cl_u +\frac 12\,\Si-\frac13\,\theta\Bigr) \,r=0 \qquad  \mbox{and} \qquad \de _\m r =0.
\end{eqnarray}
Now any first-order scalar function can be expanded as
\begin{eqnarray}
\psi = \sum_k \psi_\sch^{(k)} Q^{(k)} =\psi_\sch \, Q,
\end{eqnarray}
where $\psi_\sch$ is the scalar harmonic amplitude and the summation over $k$ is implicit in the last equality. Similarly, all vectors are expanded in terms of even ($Q_\m$) and odd ($\bar Q_\m$) parity vector harmonics which are defined respectively
\begin{eqnarray}
Q_\m =r\, \de_\m Q \qquad &&\rightarrow\qquad \de^2 Q_\m =\Bigl(K-\frac{k^2}{r^2}\Bigr) \, Q_\m \label{ever} ,\\
\bar Q_\m =r\, {\eps_\m}^\al \de_\al Q  \qquad &&\rightarrow\qquad \de^2 \bar Q_\m =\Bigl(K-\frac{k^2}{r^2}\Bigr) \, \bar Q_\m .\label{odas}
\end{eqnarray}
The vector harmonics are orthogonal ($Q^\al \bar Q_\al=0$) and they have the following properties: $\bar Q_\m = {\eps_\m}^\al Q_\al$ and $Q_\m = -{\eps_\m}^\al \bar Q_\al $. Thus any first-order vector may be expanded according to
\begin{eqnarray}
\psi_\m =\sum_k \psi_\veh ^{(k)} \, Q_\m^{(k)} +\bar \psi_\veh ^{(k)} \, \bar Q_\m^{(k)} = \psi_\veh \, Q_\m +\bar \psi_\veh \, \bar Q_\m,
\end{eqnarray}
where similarly $\psi_\veh$ and $\bar \psi_\veh$ are the vector harmonic amplitudes and the summation in the last quantity is implicit.  Also note that the 2-Laplacian acting on the vector harmonics in \eref{ever}-\eref{odas} is written in terms of the Gaussian curvature here, whereas in \cite{Betschart2004} they use a further constraint of $K=1/r^2$ which amounts to choosing a particular normalization that was convenient for their analysis.

We now present a generalization of the spherical tensor harmonics presented in  \cite{Clarkson2003} to tensor harmonics in arbitrary LRS class II space-times. The even and odd tensor harmonics are defined respectively
\begin{eqnarray}
Q_{\m\n} = r^2\, \de_{\{\m} \de_{\n\}} Q ,\qquad  &&\de^2 Q_{\m\n} = \Bigl(4\, K-\frac{k^2}{r^2}\Bigr) Q_{\m\n} ,\\
\bar Q_{\m\n} = r^2\, \eps_{\al\{\m} \de^\al \de_{\n\}} Q ,\qquad  &&\de^2 \bar Q_{\m\n} = \Bigl(4\, K-\frac{k^2}{r^2}\Bigr) \bar Q_{\m\n} ,
\end{eqnarray}
where the ``curly" brackets indicate the part that is symmetric and trace-free with respect to the 2-sheet. These are orthogonal ($Q^{\al\be} \bar Q_{\al\be}=0 $) and have the following properties: $Q_{\m\n} = {\eps_{(\m}}^\al  \bar Q_{\n)\al} $ and $\bar Q_{\m\n} = -{\eps_{(\m}}^\al Q_{\n)\al} $. Therefore, all first-order tensors may now be expanded in terms of tensor harmonics according to
\begin{eqnarray}
\psi_{\m\n} =\sum_k \psi^{(k)}_\teh \, Q^{(k)}_{\m\n} + \bar \psi^{(k)}_\teh \bar Q^{(k)}_{\m\n}= \psi_\teh \, Q_{\m\n} + \bar \psi_\teh \bar Q_{\m\n},
\end{eqnarray}
where in accord with usual terminology, $\psi_\teh$ and $\bar \psi_\teh$ are the tensor harmonic amplitudes and again the summation in the last equality is implicit.  We  also have the following relationships which also generalize those presented in \cite{Clarkson2003},
\begin{eqnarray}
\de^\al \psi_{\m\al} =  \frac r2\,\Bigl(2\,K -\frac {k^2}{r^2}\Bigr)\, \Bigl(  \psi_\teh\,Q_\m - \bar \psi_\teh \, \bar Q_\m\Bigr),\\
{\eps_{\{\m}}^\al \de^\be \psi _{\be\}\al} =  \frac r2\,\Bigl(2\,K -\frac {k^2}{r^2}\Bigr)\, \Bigl(  \bar \psi_\teh\,Q_\m + \psi_\teh \, \bar Q_\m\Bigr).
\end{eqnarray}

\section{The Gravitational and Energy-Momentum Perturbations}\label{Grperts}

We now consider both gravitational and energy-momentum perturbations to the background LRS class II space-time defined in Section \ref{backgroundlrsII}. In agreement with traditional practice we let all gravitational and energy-momentum quantities that vanish on the background LRS class II space-time simply become quantities of first-order ($\epsilon$), i.e.
\begin{eqnarray}
\mbox{first-order scalars:}     &\{\ch,\xi,\Omega,\mu, p, \cq,\Pi\}= \mc{O}(\eps),&\label{fqdas1}\\
\mbox{first-order 2-vectors:} &\{a^\m,\al^\m,\Omega^\m,\ca^\m,\Si^\m,\ce^\m,\ch^\m,\cq^\m , \Pi^{\m}\}= \mc{O}(\eps),&\label{fqdas2}\\
\mbox{first-order 2-tensors:}\,\,\,  &\{\Si_{\m\n},\zeta_{\m\n},\ce_{\m\n},\ch_{\m\n}, \Pi_{\m\n} \}= \mc{O}(\eps).&\label{fqdas3}
\end{eqnarray}
The first-order quantities given in \eref{fqdas1}-\eref{fqdas3} are all gauge-invariant under infinitesimal coordinate transformations, or more formally due to the Sachs-Stewart-Walker Lemma \cite{Sachs1964,Stewart1974}, as their corresponding background terms vanish. Furthermore, there is also the issue of choosing a particular frame in the perturbed space-time (i.e. choosing the first-order four-velocity and radial vector) as also discussed in \cite{Clarkson2003}. In general, the first-order gauge-invariant 1+1+2 quantities will not be frame invariant as they naturally depend on this choice since their underlying definitions are typically just projections and contractions with the four-velocity and radial vector. 

Now consider some perturbed quantity, $\tilde \psi$, this is expanded to first-order according to 
\begin{eqnarray}
\tilde \psi = \psi +\de \psi,
\end{eqnarray}
where $\psi$ is the corresponding background value and $\de\psi$ is the corresponding first-order part (and $\de$ is not to be confused with the covariant 2-derivative $\de_\m$).  Therefore, there are five LRS class II scalars which do not vanish on the background, and thus, they will experience first-order increments given by
\begin{eqnarray}
\{\de\ca,\de \phi,\de\Si,\de\theta,\de\ce\} = \mc{O}(\eps). \label{ngidas}
\end{eqnarray}
Furthermore, these five first-order scalars \eref{ngidas} are not gauge-invariant under the Sachs-Stewart-Walker Lemma. However, as initiated in \cite{Clarkson2003}, the 2-gradient of these scalars does vanish on the background according to  \eref{rb9}, and therefore, they become gauge-invariant quantities of first-order,
\begin{eqnarray}
\mbox{first-order 2-vectors}:\,\,\{V_\m,W_\m,X_\m,Y_\m,Z_\m\} = \mc{O}(\eps). \label{vwxyz}
\end{eqnarray}
Throughout the remainder of this paper, every equation is written in a purely gauge-invariant way. This is predominately achieved by writing everything explicitly in terms of the quantities defined in \eref{fqdas1}-\eref{fqdas2} and \eref{vwxyz}, otherwise, it is ensured that particular combinations of gauge-variant quantities are written as one combined gauge-invariant quantity.

\section{The first-order Bianchi and Ricci Identities}

The equations governing the first-order gauge-invariant 1+1+2 variables are found by decomposing the Ricci identities for both $u^\m$ and $\en^\m$, the once contracted Bianchi identities (GEM system) and the twice contracted Bianchi identities.

\subsection{Twice-contracted Bianchi Identities}

In this paper we consider the first-order energy-momentum quantities as a known source that is capable of physically perturbing the background space-time giving rise to first-order gravitational fields. Therefore, we begin with the conservation of mass equations as they will indicate how these first-order energy-momentum quantities propagate and evolve,\footnote{These are derived as follows, \eref{lum} from $ u^\al \na^\be T_{\al\be}=0$; \eref{lucq} from $\en^\al \na^\be T_{\al\be} =0$; \eref{lucjm} from $\na^\al T_{\bar \m\al}=0$.}
\begin{eqnarray}
\fl (\cl_u+\theta) \,\m +(\cl_\en +2\,\ca+\phi) \cq+ \de^\al  \cq_\al +p\,\theta +\frac 32\,\Pi\,\Si=0,\label{lum}\\
\fl\Bigl(\cl_u+\Si+\frac 43\,\theta\Bigr)\cq + (\cl_\en+\ca) p +\m\, \ca +\de^\al\Pi_\al +\Bigl(\cl_\en+\ca+\frac 32\,\phi\Bigr)\Pi =0, \label{lucq}\\
\fl (\cl_u+\theta) \cq_{\bar\m}+ (\cl_\en+\ca+\phi)\Pi_{\bar\m}+ \de_\m \Bigl(p-\frac 12\,\Pi\Bigr) +\de^\al \Pi_{\m\al}  =0 \label{lucjm}.
\end{eqnarray}.

\subsection{Gravito-electromagnetism}\label{gemchap}

The 1+1+2 GEM system is of prime importance as this paper is predominately focused on decoupling the GEM 2-tensor harmonic amplitudes. The once contracted Bianchi identities may be written in terms of the Weyl and energy-momentum tensor according to
\begin{eqnarray}
B_{\n\si\ta} := \na^\m C_{\m\n\si\ta} -  [\na_{[\si} T_{\ta]\n} +\frac 13\, g _{\n[\si} \na_{\ta]}T]=0\label{bianchids}.
\end{eqnarray} 
Before proceeding with the linearized system, we momentarily discuss the fully non-linear 1+3 GEM system, for which it is important to note that it is invariant  under the simultaneous transformation $E_{\m\n} \rightarrow H_{\m\n}$ and $H_{\m\n} \rightarrow - E_{\m\n}$ (in the absence of sources). In a recent paper \cite{Burston2007EMBP}, we used linear algebra techniques to show  that the most natural way to decouple a system with these particular invariance properties is to choose new complex dynamical variables. This has also been discussed elsewhere; for example, see \cite{Maartens1998} where they introduce a complex tensor defined $\mc{I}_{\m\n} := E_{\m\n} \pm\rmi\, H_{\m\n}$ (where $\rmi$ is the complex number). It was also this reason why we successfully decoupled the EM 2-vector harmonic amplitudes in \cite{Burston2007EMVH}.

We now turn the attention to the first-order 1+1+2 GEM  system which reduces to\footnote{These are derived as follows, \eref{lence} from $u^\al u^\be \en^\ga B_{\al\be\ga}=0$; \eref{lench} from $\eps^{\be\ga} u^\al B_{\al\be\ga} =0$; 
\eref{luce} from $  u^\al  \en^\be \en^\ga B_{\be\ga\al}=0$;  \eref{luch} from $ \eps^{\al\be} \en^\ga B_{\ga\al\be}=0$; \eref{lencem} from $ u^\be u^\ga B_{\bar\m\be\ga}=0$; \eref{lenchm}  from ${\eps_{\bar\m}}^{\be\ga} u^\al  B_{\al\be\ga}=0$; \eref{lucem} from $\en^\n u^\ga B_{(\bar\m\n)\ga} =0$; \eref{luchm} from $\en^\n \,{\eps_{(\bar\m}}^{\al\be} B_{\n)\al\be}=0$; \eref{lucemn} from $u^\al B_{(\bar\m\bar\n)\al}=0$;  \eref{luchmn} from ${\epsilon_{(\bar\m}}^{\al\be} B_{\bar\n)\al\be}=0$.} 

\begin{eqnarray}
\fl\de\Bigl[\Bigl(\cl_\en+\frac 32\,\phi\Bigr)\ce\Bigr]+\de^\al \ce_\al =\Re[\cg],\label{lence}\\
\fl\Bigl(\cl_\en+\frac 32\,\phi\Bigr)\ch+\de^\al \ch_\al +3\,\ce\,\Omega=\Im[\cg] \label{lench},\\
\fl\de\Bigl[\Bigl(\cl_u-\frac32\,\Si+\theta\Bigr)\ce\Bigr]-\epsilon^{\al\be}\,\de_\al \mc{H}_\be=\Re[\cf],\label{luce}\\
\fl\Bigl(\cl_u-\frac32\,\Si+ \theta\Bigr)\ch+\epsilon^{\al\be}\,\de_\al \ce_\be +3\, \ce\,\xi= \Im[\cf]\label{luch},\\
\fl (\cl_\en+\phi)\,\ce_{\bar\m}+\de^\al \,\ce_{\m\al}  -\,\frac12\,X_\m+\frac32 \, \Sigma\, {\epsilon_\m}^\al\mc{H}_\al\,+\frac 32 \,\ce\,a_\m  =\Re [\cg_\m],\label{lencem}\\
\fl (\cl_\en+\phi)\ch_{\bar\m}+\de^\al \ch_{\m\al} -\frac12\,\de_\m \ch -\frac 32\,\Si\,\,{\epsilon_\m}^\al\,\ce_\al\,+\frac 32\,\ce\,{\epsilon_\m}^\al  \,(\Si_\al +{\eps_\al}^\be\Omega_\be ) =\Im[\cg_\m],\label{lenchm}\\
\fl\Bigl(\cl_u-\Si+\frac 23 \theta\Bigr) \ce_{\bar\m}- {\epsilon_{\m}}^\al \de^\be \ch_{\al\be}- \frac 12{\epsilon_\m}^\al\left[ \de_\al\ch-(2\ca-\phi)\ch_\al  \right]+\frac 32 \ce\,\al_\m=\Re[\cf_\m]\label{lucem},\\
\fl\Bigl(\cl_u-\Si+\frac 23 \theta\Bigr) \ch_{\bar\m}+ {\epsilon_{\m}}^\al \de^\be \ce_{\al\be}+ \frac 12{\epsilon_\m}^\al\,\left[ X_\al-(2\ca-\phi)\ce_\al \right]+\frac 32 \ce\,{\eps_\m}^\al\ca_\al=\Im[\cf_\m]\label{luchm},\\
\fl\Bigl(\cl_u +\frac52\Si+\frac13\theta\Bigr)\ce_{\bar \m\bar\n} + {\eps_{(\m}}^\al\left(\cl_\en +2\ca-\frac 12 \phi\right)\ch_{\n)\al}  -{\eps_{\{\m}}^\al \de_{|\al|}\ch_{\n\}}  +\frac 32 \ce\Si_{\m\n} =\Re[\cf_{\m\n}],\label{lucemn}\\
\fl\Bigl(\cl_u +\frac 52\Si+\frac13  \theta\Bigr)\ch_{\bar \m\bar\n} -{\eps_{(\m}}^\al \left(\cl_\en +2\ca-\frac 12\phi\right)\ce_{\n)\al}  +{\eps_{\{\m}}^\al \de_{|\al|}\ce_{\n\}}+\frac 32\ce {\eps_{(\m}}^\al\zeta_{\n)\al} =\Im[\cf_{\m\n}] \label{luchmn}.\nn\\
\end{eqnarray}
The first-order energy-momentum source terms have been suitably defined in a complex form for later convenience as
\begin{eqnarray}
\fl \cf :=-\frac 12\,(\m+p)\Si -\frac 13\,\Bigl(\cl_\en+2\,\ca-\frac 12\,\phi\Bigr) \cq +\frac 16\,\de^\al\cq_\al  -\frac 12\,\Bigl(\cl_u+\frac 12\,\Si+\frac 13\,\theta\Bigr)\Pi \nn\\
+\rmi \,\frac12\,\eps^{\al\be} \de_\al \Pi_\be,  \label{cf}\\
\fl\cg:= \frac13\, \cl_\en \mu+\frac 12\, \cq\,\Bigl(\Si-\frac23 \,\theta\Bigr)-\frac 12\, \de^\al \Pi_\al -\frac 12\, \Bigl(\cl_\en+\frac 32\,\phi\Bigr)\Pi -\rmi\,\frac12 \, \epsilon^{\al\be} \de_\al \cq_\be, \\
\fl \cf_\m := -\frac 12\,\left[ \cl_u \Pi_{\bar\m}      + \Bigl(\ca -\frac 12\,\phi\Bigr)\cq_{\bar \m} + \de_\m \cq \right]\nn\\
+\rmi\, \frac12\,{\epsilon_\m}^\al\left[\frac13\, \de_\al (\m+3\,\Pi)- \Bigl(\Si+\frac 13\,\theta\Bigr)\, \cq_\al -\Bigl(\cl_\en+\frac 12\,\phi\Bigr)\Pi_\al \right],\\
\fl\cg_\m :=\frac13\, \de_\m \Bigl(\m+\frac 34\,\Pi\Bigr)-\frac 14\, \Bigl(\Si+\frac 43\,\theta\Bigr)\, \cq_\m 
-\frac 12\, (\cl_\en+\phi)\Pi_{\bar\m} -\frac 12\,\de^\al \Pi_{\m\al}  \nn\\
+ \rmi\,\frac 12\,{\eps_\m}^\al\,\Bigl( \cl_\en \cq_\al - \de_\al \cq +\frac 32 \,\Si\,\Pi_\al \Bigr),\\
\fl\cf_{\m\n} :=- \frac12 \de_{\{\m} \cq_{\n\}}        - \frac12\left(\cl_u+\frac 12\,\Si-\frac 13\,\theta\right) \Pi_{\bar\m\bar\n} \nn\\
+\rmi\,\frac 12\,\Bigl[ {\eps_{\{\m}}^\al \de_{|\al|} \Pi_{\n\}}-  {\eps_{(\m}}^\al\left(\cl_\en-\frac 12\,\phi\right)\Pi_{\bar\n)\al}   \Bigr]. \label{cfmn}
\end{eqnarray}
The first-order GEM system \eref{lence}-\eref{luchmn} generalize those given in \cite{Clarkson2003} in two significant ways; they generalize from the Schwarzschild perturbations towards an arbitrary vacuum LRS class II space-time and they also generalize from the vacuum energy-momentum perturbations towards a full energy-momentum perturbation.  Furthermore, a very recent independent study of these equations for LRS space-times has been carried out in \cite{Clarkson2007}. We have also taken  a lot of care to ensure that all quantities are gauge-invariant; for example,  the first-order term in \eref{lence}, $\de[(\cl_\en+\frac 32\,\phi)\ce]$, is gauge-invariant as its corresponding background term vanishes according to \eref{rb3}, i.e. $(\cl_\en+\frac 32\,\phi)\ce=0$. However, we now choose to rewrite \eref{lence}-\eref{luch} in terms of the 2-gradient quantity $X_\m$ defined in \eref{xysa}. Thus, new complex variables are chosen according to the invariance properties of the 1+3 GEM system discussed above and, without loss of generality, we write the GEM system in a new 1+1+2 complex form,
\begin{eqnarray}
\fl \Bigl(\cl_\en+\frac 32\phi\Bigr) \ex_\m  + \de_\m \de^\al \Phi_\al +\frac 32\ce\Bigl[Y_\m -\phi a_\m-2\,\Bigl(\Si-\frac 23\theta\Bigr){\eps_\m}^\al \Omega_\al+\rmi2\de_\m \Omega \Bigr]  = \de_\m \cg,\label{lenxe}\\
\fl \Bigl(\cl_u-\frac 32\,\Si+\theta\Bigr) \ex_{\bar\m}+\rmi\,\de_\m (\eps^{\al\be} \de_\al \Phi_\be)\nn\\
 -\frac32\,\ce\, \left[\ca_\m\,\Bigl(\Si-\frac 23\,\theta\Bigr)+\phi\,(\Si_\m-{\eps_\m}^\al \Omega_\al+\al_\m) +W_\m-\rmi\,2\, \de_\m \xi \right]=\de_\m\cf,\\
\fl(\cl_\en+\phi)\,\Phi_{\bar\m}+\de^\al \,\Phi_{\m\al}  -\,\frac12\,\de(\de_\m\, \Phi)-\rmi\,\frac32 \, \Sigma\, {\epsilon_\m}^\al\Phi_\al\,+\frac 32 \,\ce\,\Lambda_\m=\cg_\m,\label{lenphim}\\
\fl\Bigl(\cl_u-\Si+\frac 23\, \theta\Bigr) \Phi_{\bar\m}+ \rmi\,{\epsilon_\m}^\al \de^\be \Phi_{\al\be}+\rmi\, \frac 12\,{\epsilon_\m}^\al\,\left[ \ex_\al-(2\,\ca-\phi)\,\Phi_\al  \right]+\frac 32 \,\ce\,\Upsilon_\m=\cf_\m, \label{luphimdef}\\
\fl\Bigl(\cl_u +\frac52\Si+\frac13\theta\Bigr)\Phi_{\bar \m\bar\n}-\rmi{\epsilon_{(\m}}^\al \Bigl(\cl_\en +2\ca-\frac 12\, \phi\Bigr)\Phi_{\n)\al}  +\rmi\,{\epsilon_{\{\m}}^\al \de_{|\al|}\Phi_{\n\}}+\frac 32 \,\ce\,\Lambda_{\m\n} = \cf_{\m\n} \label{lnphimn},
\end{eqnarray} 
where  
\begin{eqnarray}
\fl \ex_\m := X_\m +\rmi \, \de_\m\ch, \qquad\Phi_\m := \ce_\m +\rmi \,\ch_\m \qquad\mbox{and} \qquad \Phi_{\m\n}:= \ce_{\m\n} + \rmi \, \ch_{\m\n}. \label{defpohids}
\end{eqnarray}
\footnote{It is also possible to choose the complex conjugates, i.e. $\Phi_{\m\n}^*$, $\Phi_\m^*$ and $\Phi^*$ and the corresponding governing equations are simplify found by taking the complex conjugate of the equations governing $\Phi_{\m\n}$, $\Phi_\m$ and $\Phi$.} Furthermore, whilst constructing these complex equations, several other terms naturally combine and therefore, 3  new complex definitions are
\begin{eqnarray}
\fl\Upsilon_\m := \al_\m +\rmi\,{\epsilon_\m}^\al \ca_\al,\,\,\,\,\,
\Lambda_\m  :=  a_\m  +\rmi\,{\epsilon_\m}^\al(  \Si_\al+ {\epsilon_\al}^\be\Omega_\be ) \,\,\,\,\,\mbox{and}\,\,\,\,\,\,
\Lambda_{\m\n} := \Sigma_{\m\n}+\rmi \,{\epsilon_{(\m}}^\al \, \zeta_{\n)\al}.\label{newcomasd}
\end{eqnarray}
In Section \ref{decouplingphimn} we will use the complex GEM system \eref{lenxe}-\eref{lnphimn} to fully decouple the complex GEM 2-tensor, $\Phi_{\m\n}$, from all the remaining 1+1+2 quantities.

\subsection{The 1+1+2 Ricci Identities}

The Ricci identities for both $u^\m$ and $\en^\m$ are defined conveniently as
\begin{eqnarray}
Q_{\m\n\si} := 2\,\na_{[\m} \na_{\n]} u_\si - R_{\m\n\si\ta} u^\ta =0,\label{ricciforlittlenm}\\
R_{\m\n\si} := 2 \na_{[\m} \na_{\n]} \en_\si - R_{\m\n\si\ta} \en^\ta=0\label{riccifornm},
\end{eqnarray}
where $R_{\m\n\si\ta}$ is the Riemann tensor. We now linearize these, reduce them to 1+1+2 form and categorize them into constraint, propagation, transportation  and evolution equations. We also make two new definitions for combinations that arise quite frequently,
\begin{eqnarray}
\lambda_\m := \Si_\m - {\eps_\m}^\al \Omega_\m \qquad\mbox{and}\qquad \ups_\m := \Si_\m + {\eps_\m}^\al \Omega_\m,
\end{eqnarray}
such that the following system can be written in a more readable form.

\begin{itemize}
 \item Constraint equations\footnote{ \eref{constratforw} from a combination of $\en^\m u^\si R_{\m\bar\n\si}=0$, $\cs^{\m\si} Q_{\m\bar\n\si} =0 $ and $ \en^\m \en^\si Q_{\m\bar\n\si}=0$; \eref{conforym} from $\cs^{\n\si} R_{\bar\m\n\si}=0$ and  \eref{divlam} from $\eps^{\m\n} u^\si R_{\m\n\si}=0$. }
\begin{eqnarray}
 \fl   \dst_\m +\phi\,\lam_\m  +2\,\de^\al \Si_{\m\al} +2\,{\eps_\m}^\al \ch_\al  +2\,{\epsilon_\m}^\al \de_\al\Omega =-\cq_\m \label{constratforw},\\
\fl Y_\m -2\,{\epsilon_\m}^\al \de_\al\xi -2\,\de^\al\zeta_{\m\al} +2\,\ce_\m +\Bigl(\Si-\frac 23\, \theta\Bigr)\lam_\m=- \Pi_\m \label{conforym},\\
\fl \eps^{\al\be}\de_\al \lambda_\be -(2\,\ca-\phi) \Omega +3\,\xi\,\Si-\ch=0\label{divlam}.
\end{eqnarray}
  \item Propagation equations\footnote{\eref{lenphi} from  $\en^\m \cs^{\n\si} R_{\m\n\si} =0$; \eref{lensimthe} from $\en^\m\cs^{\n\si} Q_{\m\n\si} =0$; \eref{lenxi} from  $\en^\m \eps^{\n\si}  R_{\m\n\si}=0$
; \eref{lenomega} from $\eps^{\m\n\si} Q_{\m\n\si}=0$; \eref{lensiminomem} from $D^\al \si_{\m\al}$ equation and $\en^\m u^\si R_{\m\bar\n\si}=0$; \eref{lensimn} from $ \en^\m Q_{\m(\bar\n\bar\si)}=0$; \eref{lenzetamn} from   $\en^\m R_{\m(\bar\n\bar\si)} =0$.}
 \begin{eqnarray}
\fl \de\Bigl\{\Bigl(\cl_\en+\frac 12\,\phi\Bigr)\phi+\Bigl(\Si+\frac 13\,\theta\Bigr)\Bigl(\Si-\frac 23\,\theta\Bigr) +\ce\Bigr\}-\de^\al a_\al =-\frac 23\,\m-\frac 12\,\Pi \label{lenphi},\\
\fl \de\Bigl\{ \cl_\en \Bigl(\Si-\frac 23\,\theta\Bigr) +\frac 32\,\phi\,\Si \Bigr\}+\de^\al \ups_\al  = -\cq \label{lensimthe},\\
\fl (\cl_\en+\phi) \xi -\Bigl(\Si+\frac 13\,\theta\Bigr)\Omega -\frac 12\,\eps^{\al\be} \de_\al a_\be =0,\label{lenxi}\\
\fl (\cl_\en-\ca+\phi)\Omega+\de^\al  \Omega_\al=0, \label{lenomega}\\
\fl\cl_\en \lam_{\bar\m} +\frac 12\,\phi\,\ups_\m -2\,\ca{\eps_\m}^\al \Omega_\al -\de_\m\Bigl(\Si+\frac 13\,\theta\Bigr)+\frac 32\,\Si \,a_\m -{\eps_\m}^\al \ch_\al   =-\frac 12\,\cq_\m ,\label{lensiminomem}\\
 \fl \Bigl(\cl_\en-\frac 12\,\phi\Bigr) \Si_{\bar\m\bar\n} -\frac 32\,\Si\,\zeta_{\m\n} -{\epsilon_{(\m}}^\al \ch_{\n)\al} -\de_{\{\m}\ups_{\n\}} =0,\label{lensimn}\\
\fl \cl_\en \zeta_{\bar\m\bar\n} -\Bigl(\Si+\frac 13\,\theta\Bigr) \Si_{\m\n} +\ce_{\m\n} -\de_{\{\m} a_{\n\}}  =-\frac 12\,\Pi_{\m\n}. \label{lenzetamn}
\end{eqnarray}
  \item Transportation\footnote{\eref{lusidas} from $u^\m \en^\n u^\si R_{\m\n\si}$; \eref{clcam-lusiom} from $\en^\m u^\n {\cs_\si}^\ga Q_{\m\n\ga} =0$;  \eref{lua-lenalpms} from $u^\al \en^\be R_{\al\be\bar\m}=0$.}
\begin{eqnarray}
 \fl \de\Bigl\{\Bigl(\cl_u+\Si+\frac 13\,\theta\Bigr)\Bigl(\Si+\frac 13\,\theta\Bigr) -(\cl_\en+\ca)\ca+\ce\Bigr\}    =-\frac 16(\m+3\, p-3\,\Pi) \label{lusidas},\\
\fl\Bigl(\cl_u+\Si+\frac 13\,\theta\Bigr)\ups_{\bar\m}-\Bigl(\cl_\en+\ca-\frac 12\,\phi\Bigr) \ca_{\bar\m} -\ca \, a_\m +\frac 32\,\Si \,\al_\m +\ce_\m =\frac 12\,\Pi_\m,\label{clcam-lusiom}\\
\fl\Bigl(\cl_u+\frac 32 \Si\Bigr) a_{\bar\m} \hspace{-0.1cm}-\hspace{-0.1cm}(\cl_\en+\ca) \alpha_{\bar\m} -\Bigl(\ca-\frac 12\phi\Bigr) \ups_\m \hspace{-0.1cm}+\hspace{-0.1cm}\Bigl(\Si+\frac 13 \theta\Bigr) \ca_\m -{\epsilon_\m}^\al \ch_\al =-\frac 12\cq_\m \label{lua-lenalpms} .
\end{eqnarray}
 \item Evolution equations\footnote{\eref{luphi} from $u^\m \,\cs^{\n\si} R_{\m\n\si} =0 $; \eref{lusimthe} from   $u^\m \cs^{\n\si} Q_{\m\n\si}=0$; \eref{luxi} from $u^\m \eps^{\n\si}  R_{\m\n\si}=0$; \eref{luomega} from $u^\m \eps^{\n\si} Q_{\m\n\si}=0$; \eref{lusi-om} from  $u^\m \en^\si {\cs_\n}^\al Q_{\m\al\si}=0$ ;\eref{luzetamn} from $u^\m R_{\m(\bar\n\bar\si)}=0$; \eref{lusimn} from $u^\m Q_{\m(\bar\n\bar\si)}=0$.}
  \begin{eqnarray}
\fl  \de\Bigl\{\Bigl(\cl_u-\frac 12\,\Si+\frac 13\,\theta\Bigr)\phi +\ca\,\Bigl(\Si-\frac 23\,\theta\Bigr)\Bigr\}  - \de^\ga  \al_\ga =\cq \label{luphi},\\
\fl \de\Bigl\{\Bigl(\cl_u-\frac 12\, \Si+\frac 13\,\theta\Bigr)\Bigl(\Si-\frac 23\,\theta\Bigr)+\ca\,\phi  +\ce\Bigr\} +\de^\al  \ca_\al= \frac 13\Bigl(\m+3\,p+\frac 32\, \Pi\Bigr)\label{lusimthe} ,\\
\fl \Bigl(\cl_u -\frac 12\,\Si +\frac 13\,\theta\Bigr) \xi -\frac 12\, \eps^{\al\be}\de_\al \al_\be  -\Bigl(\ca-\frac 12\,\phi\Bigr) \Omega -\frac 12\,\ch =0, \label{luxi} \\
\fl \Bigl(\cl_u -\Si-\frac 23\,\theta \Bigr) \Omega-\ca\,\xi - \frac12\,\eps^{\al\be} \de_\al \ca_\be=0,\label{luomega}\\
  \fl (\cl_u +\theta)\lam_{\bar \m} -Z_\m -\Bigl(\ca-\frac 12\,\phi\Bigr)\ca_\m +\frac 32\,\Si\, \al_\m +\ce_\m =\frac 12\,\Pi_\m ,\label{lusi-om}\\
\fl \Bigl(\cl_u +\frac 12\,\Si-\frac 13\,\theta\Bigr) \zeta_{\bar\m\bar\n}-\Bigl(\ca-\frac 12\,\phi\Bigr) \Si_{\m\n} -{\epsilon_{(\m}}^\al \ch_{\n)\al} -\de_{\{\m}\al_{\n\}} =0 ,\label{luzetamn}\\
\fl  \cl_u \Si_{\bar\m\bar\n} -\ca\,\zeta_{\m\n} - \de_{\{\m}\ca_{\n\}} +\ce_{\m\n}=\frac 12\,\Pi_{\m\n} \label{lusimn}.
\end{eqnarray}
\end{itemize}
Similarly, these 1+1+2 Ricci identities \eref{constratforw}-\eref{lusimn} are again a significant generalization of  the results in \cite{Clarkson2003}. They now include full energy-momentum sources and moreover, they  are for arbitrary vacuum LRS class II space-times.  Moreover, the very recent independent study by Clarkson \cite{Clarkson2007} presents the equations for LRS space-times. For the subsequent decoupling of the complex GEM 2-tensor, we require evolution, transportation and propagation equations for the complex variables defined in \eref{newcomasd}\footnote{\eref{ldjsavcjnkxc} from \eref{clcam-lusiom} and \eref{lua-lenalpms}; \eref{lulambdamu} from \eref{luzetamn}, \eref{lusimn};  \eref{lenlambdamu} from \eref{lensimn} and \eref{lenzetamn}}
\begin{eqnarray}
\fl \Bigl(\cl_u+\frac 32\, \Si\Bigr) \Lambda_{\bar\m} -(\cl_\en+\ca) \Upsilon_{\bar\m}+\rmi {\eps_\m}^\al \Phi_\al- \rmi\,\ca {\eps_\m}^\al \Lambda_\al + \frac 12\,\phi\, (\ups_\m+\rmi{\eps_\m}^\al \ca_\al) \nn\\
\fl+\Bigl(\Si+\frac 13 \theta\Bigr) \ca_\m-\rmi\,\frac 12 \,\Bigl(\Si-\frac 23\theta\Bigr) {\eps_\m}^\al\,  \ups_{\al}
  +\rmi \,\frac 32\,\Si\, {\eps_\m}^\al \al_\al=-\frac 12(\cq_\m -\rmi \,{\eps_\m}^\al \Pi_\al ), \label{ldjsavcjnkxc}\\
\fl\cl_u \Lambda_{\bar\m\bar\n} +\Phi_{\m\n} -\rmi\,\ca\,{\epsilon_{(\m}}^\al \Lambda_{\n)\al} +\rmi\,\frac 12 \,\phi\,{\epsilon_{(\m}}^\al \Si_{\n)\al} \nn\\
+\rmi\,\frac12\,\Bigl(\Si-\frac 23\,\theta\Bigr) {\epsilon_{(\m}}^\al \zeta_{\n)\al} -\rmi\,{\epsilon_{\{\m}}^\al \de_{\n\}}\Upsilon_\al =\frac 12\,\Pi_{\m\n} ,\label{lulambdamu}\\
\fl\cl_\en \Lambda_{\bar\m\bar\n} + \rmi {\epsilon_{(\m}}^\al \Phi_{\n)\al}-\rmi\Bigl(\Si+\frac 13\,\theta\Bigr) {\epsilon_{(\m}}^\al \Si_{\n)\al}-\frac 32\,\Si\,\zeta_{\m\n}  \nn\\
 -\frac 12\,\phi\,\Si_{\m\n} -\rmi\, {\epsilon_{\{\m}}^\al \de_{\n\}} \Lambda_\al  =-\rmi\,\frac12 {\epsilon_{(\m}}^\al \Pi_{\n)\al}\label{lenlambdamu}.
\end{eqnarray}

\subsection{Commutation relationships}

Finally, we present how the various derivatives defined in this paper commute and generalize the results from \cite{Betschart2004}, 

\begin{eqnarray}
\Bigl(\cl_u+\Si+\frac 13\,\theta\Bigr) \cl_\en \Phi_{\bar\m\dots\bar\n} - (\cl_\en+\ca) \cl_u \Phi_{\bar\m\dots\bar\n} =0, \label{com1}\\
\cl_u \,\de_\si \Phi_{\bar\m\dots\bar\n} - \de_\si \,\cl_u \Phi_{\bar\m\dots\bar\n}=0 \label{com2},\\
\cl_\en \,\de_\si \Phi_{\bar\m\dots\bar\n} - \de_\si \,\cl_\en \Phi_{\bar\m\dots\bar\n} =0. \label{com3}
\end{eqnarray}
where $\Phi_{\m\dots\n}$ represents a first-order scalar, first-order 2-vector and a first-order 2-tensor. The commutators not only play a vital role in decoupling the equations at hand, they also provide a rigorous test that the equations present here are correct and accurate. Every equation \eref{cf}-\eref{lnphimn} and \eref{constratforw}-\eref{lenlambdamu} has been checked to satisfy all of the commutator relationships \eref{com1}-\eref{com2} and this is inclusive of careful checks of all energy-momentum source terms \eref{cf}-\eref{cfmn}.

\section{Decoupling the complex GEM 2-tensor and its tensor harmonic amplitudes}\label{decouplingphimn}

We use the complex 1+1+2 Bianchi identities  \eref{lenxe}-\eref{lnphimn} to construct a new, covariant and gauge-invariant, equation governing the first-order complex GEM 2-tensor $\Phi_{\m\n}$. This is with a complete description of the, covariant and gauge-invariant, first-order energy-momentum sources. It begins by taking the Lie derivative with respect to $u^\m$ of \eref{lnphimn}. It is then required to use the commutation relationships \eref{com1}-\eref{com2} followed by substitutions of \eref{lenphim} through to \eref{lnphimn}. Finally, \eref{lulambdamu} and \eref{lenlambdamu} are used for further simplifications to obtain
\begin{eqnarray}
\fl[(\cl_u+\theta) \cl_u -(\cl_\en+\ca+\phi) \cl_\en -V ]\Phi_{\m\n} \nn\\
-\rmi\,{\eps_{(\m}}^\al\left[(4\,\ca-2\,\phi)\cl_u -6\,\Si\,  \cl_\en + U \right]\Phi_{\n)\al} =\cm_{\m\n}. \label{bpequation}
\end{eqnarray}
The two background scalars related to the potential, and the first-order energy-momentum source, have been defined respectively
\begin{eqnarray}
\fl V := \de^2+8\,\ce-4\,\ca^2+4\,\ca\,\phi-\phi^2+9\,\Si^2-3\,\Lambda,\\
\fl U:=2\,\Bigl(\cl_u-\Si+\frac 23\,\theta\Bigr) \ca -3\,\Bigl(\cl_\en+\frac 76\,\phi\Bigr) \Si -\frac23 \,\theta\,\phi-2\,\Lambda,\\
\fl\cm_{\m\n}:= \Bigl(\cl_u -\frac 52\Si+\frac23\theta\Bigr)\cf_{\bar\m\bar\n} +\rmi\, {\epsilon_{(\m}}^\al \Bigl(\cl_\en-\ca+\frac 32 \phi\Bigr) \cf_{\n)\al} -\rmi{\epsilon_{\{\m}}^\al \de_{|\al|} \cf_{\n\}} -\de_{\{\m}  \cg_{\n\}}.
\end{eqnarray}
It was possible to eliminate all Lie derivatives in  $V$ and write it explicitly as algebraic combinations of the background LRS class II scalars . However, the Lie derivatives in the other potential term, $U$, cannot be reduced any further because there is no evolution equation for $\ca$.  

Thus \eref{bpequation} demonstrates that, for arbitrary vacuum LRS class II space-times, the complex GEM 2-tensor decouples from the remaining GEM and 1+1+2 quantities. We next show how this 2-tensor decouples further by using a tensor harmonic expansion, but we first take a closer inspection of the energy-momentum source, $\cm_{\m\n}$,
\begin{eqnarray}
\fl\cm_{\m\n} = -\frac 12 \Bigl\{  \Bigl(\cl_u -2\,\Si +\frac 13\,\theta\Bigr) \cl_u \Pi_{\m\n} +(\cl_\en-\ca+\phi) \cl_\en \Pi_{\m\n} -\cm\, \Pi_{\m\n}- 2\, \de_{\{\m} \de^\al \Pi_{\n\}\al}\Bigr\}\nn\\
\fl +2\,(\cl_\en +\phi) \de_{\{\m} \Pi_{\n\}} +2\,\Bigl(\Si+\frac13\,\theta\Bigr) \de_{\{\m} \cq_{\n\}} +\frac 12 \de_{\{\m} \de_{\n\}} (p+2\,\Pi), \nn\\
\fl+\rmi\, {\eps_{\{\m}}^\al \Bigl\{-\Bigl(\cl_u-\frac 12\,\Si+\frac 13\,\theta\Bigr) \cl_\en \Pi_{\n\} \al} +\Bigl(\ca-\frac 12\,\phi\Bigr)\, \cl_u \Pi_{\n\}\al}   +  \phi\,\Bigl(\Si-\frac 23\,\theta\Bigr)\, \Pi_{\n\}\al} \nn\\
\fl+\Bigl(\cl_u-2\,\Si+\frac 13\,\theta\Bigr)  \de_{\n\}} \Pi_\al  -(\cl_\en-\ca+\phi) \de_{\n\}} \cq_\al + \de_{\n\}}\de_\al \cq\Bigr\},
\end{eqnarray}
where 
\begin{eqnarray}
\cm = \frac 12 \Bigl(\Si-\frac 23\,\theta\Bigr)^2 +\frac 12\,\ca\,\phi +\frac 12\,\phi^2-\frac 12\,\ce.
\end{eqnarray}
It is interesting to see which energy-momentum terms play an important role in the evolution and propagation of the complex GEM 2-tensor. By considering the ``principle part", or the parts which involve second-order Lie derivatives, it seems that the first-order anisotropic stress may have a predominate influence here.

\subsection{Decoupling the complex GEM 2-tensor harmonic amplitudes}

The complex GEM  tensor, $\Phi_{\m\n}$, and the energy-momentum source, $\cm_{\m\n}$, are expanded using tensor harmonics according to
\begin{eqnarray}
\Phi_{\m\n} =\Phi_\teh \, Q_{\m\n} + \bar \Phi_\teh\, \bar Q_{\m\n} \qquad\mbox{and}\qquad\cm_{\m\n} =\cm_\teh \, Q_{\m\n} + \bar \cm_\teh\, \bar Q_{\m\n}.\nn
\end{eqnarray}
Consequently, \eref{bpequation} results in two coupled equations of the form
\begin{eqnarray}
\fl\Bigl[\Bigl(\cl_u-2\,\Si+\frac 73\,\theta\Bigr)\cl_u-(\cl_\en+\ca+3\,\phi) \cl_\en -\tilde V \Bigr]\Phi_\teh \nn\\
+\rmi\, \Bigl[6\,\Si\,\cl_\en-( 4\,\ca-2\,\phi) \cl_u - \tilde U \Bigr]\bar \Phi_\teh =\cm_\teh,\label{waveforphit}\\
\fl\Bigl[\Bigl(\cl_u-2\,\Si+\frac 73\,\theta\Bigr)\cl_u-(\cl_\en+\ca+3\,\phi) \cl_\en -\tilde V \Bigr]\bar\Phi_\teh \nn\\
-\rmi\, \Bigl[6\,\Si\,\cl_\en-( 4\,\ca-2\,\phi) \cl_u - \tilde U \Bigr] \Phi_\teh =\bar\cm_\teh,\label{waveforbarphit}
\end{eqnarray}
where new potential terms are defined
\begin{eqnarray}
\tilde V := -\frac{k^2}{r^2}+2\,\ce-4\,\ca^2+4\,\ca\,\phi +\frac 32\,\phi^2 +\frac{13}2 \,\Si^2 -\frac {10}9\,\theta^2+\frac{10}3\,\Si\,\theta ,\nn\\
\tilde U :=  2\,\Bigl(\cl_u-3\,\Si +2\,\theta\Bigr)\ca - 3 \,\Bigl(\cl_\en+\frac 52\,\phi\Bigr) \Si -2\,\theta\,\phi.
\end{eqnarray}
By inspecting the coupled system \eref{waveforphit} and \eref{waveforbarphit}, it is clear that they are invariant under the simultaneous transformation of $\Phi_\teh \rightarrow \bar\Phi_\teh$ and $\bar\Phi_\teh \rightarrow -\Phi_\teh$, and similarly for the sources, $\cm_\teh \rightarrow\bar\cm_\teh$ and $\bar \cm_\teh\rightarrow-\cm_\teh$. Thus, the coupled system \eref{waveforphit}-\eref{waveforbarphit} is precisely of the form  as discussed at the beginning of Section \ref{gemchap}. Therefore, they will decouple quite naturally by constructing two new complex dependent variables,
\begin{eqnarray}
\Phi_+ := \Phi_\teh+ \rmi\,\bar\Phi_\teh\qquad\mbox{ and }\qquad\Phi_- := \Phi_\teh- \rmi\,\bar\Phi_\teh. \label{Phiphimds}
\end{eqnarray}
We also define a new complex energy-momentum source $\cm_\pm := \cm_\teh \pm \rmi\,\bar\cm_\teh$ and potential $V_\pm :=\tilde V \pm \tilde U$, where the ``$\pm$" is relative. Therefore, by taking complex combinations of  \eref{waveforphit} and \eref{waveforbarphit}, we find two new decoupled equations given by
\begin{eqnarray}
\fl\Bigl\{ \Bigl[\cl_u -2\,\Si+\frac 73\, \theta +(2\,\phi-4\,\ca)\Bigr]\cl_u -\left(\cl_\en+\ca+3\,\phi-6\,\Si\right)\cl_\en - V_+\Bigr\} \Phi_+ = \cm_+,\label{avephi+}\\
\fl\Bigl\{ \Bigl[\cl_u -2\,\Si+\frac 73\, \theta -(2\,\phi-4\,\ca)\Bigr]\cl_u -\left(\cl_\en+\ca+3\,\phi+6\,\Si\right)\cl_\en - V_-\Bigr\} \Phi_- = \cm_-.\label{avephi-}
\end{eqnarray}
It is vital to point out here that, since the covariant differential operators in \eref{avephi+}-\eref{avephi-} are purely real, by taking the real and imaginary parts separately there are actually four {\it real} decoupled quantities. It is now of interest to see how  $\Phi_\pm$ relates back to the real GEM 2-tensor harmonic amplitudes. The GEM 2-tensors are expanded according to
\begin{eqnarray}
\ce_{\m\n} = \ce_\teh\, Q_{\m\n} +\bar\ce_\teh\,\bar Q_{\m\n} \qquad\mbox{and}\qquad \ch_{\m\n} = \ch_\teh\, Q_{\m\n} +\bar\ch_\teh\,\bar Q_{\m\n}.
\end{eqnarray}
Here, the polar perturbations are $\ce_\teh$ and $\bar\ch_\teh$ whereas the axial perturbations are  $\bar \ce_\teh$ and $\ch_\teh$. Moreover,  a full categorization of all the harmonic amplitudes of the 1+1+2 dependent variables into polar and axial perturbations is presented in \cite{Clarkson2003}. The definition  \eref{defpohids} now implies,
\begin{eqnarray}
\Phi_\teh: = \ce_\teh +\rmi \, \ch_\teh \qquad \mbox{and}\qquad \bar \Phi_T := \bar\ce_\teh+\rmi \, \bar \ch_\teh,
\end{eqnarray}
and by subsequently using \eref{Phiphimds} we find,
\begin{eqnarray}
\fl \Phi_+ = (\ce_\teh-\bar{ \ch}_\teh) +\rmi \, (\bar \ce_\teh +\ch_\teh) \qquad\mbox{and}\qquad\Phi_- = (\ce_\teh+\bar{ \ch}_\teh) -\rmi \, (\bar \ce_\teh -\ch_\teh).
\end{eqnarray}
Thus, the four precise combinations of the four real GEM 2-tensor harmonic amplitudes  which decouple are,
\begin{eqnarray}
\mbox{Decoupled polar perturbations:     }\{\ce_\teh + \bar\ch_\teh,\ce_\teh- \bar\ch_\teh\} ,\label{dasD}\\
\mbox{Decoupled axial perturbations:     }\{\ch_\teh  + \bar\ce_\teh,\ch_\teh- \bar\ce_\teh\} .\label{das}
\end{eqnarray}
Moreover, it is clear that if the 4 decoupled quantities are known, then simple linear combinations will reveal each of $\ce_\teh$, $\bar\ch_\teh$, $\ch_\teh$ and $\bar\ce_\teh$.

\section{Summary}

This paper discussed covariant and gauge-invariant  gravitational and energy-momentum perturbations on arbitrary vacuum LRS class II space-times. We showed how particular combinations of the first-order GEM quantities decouple at two different levels. The first was a complex tensorial equation governing the complex GEM 2-tensor $\Phi_{\m\n}$ \eref{bpequation}. The second involved a tensor harmonic expansion of the GEM 2-tensors and resulted in four {\it real} equations \eref{avephi+}-\eref{avephi-}. Of particular interest is that we have found the precise combinations of the GEM 2-tensor harmonic amplitudes that decouple, and these were separated out into polar and axial perturbations according to \eref{dasD}-\eref{das}.

\end{document}